\documentclass[secnumarabic,amssymb, nobibnotes, aps, prd]{revtex4-1}

\usepackage{graphicx} 

\begin{document}
\title{Efficient fluorescence collection and ion imaging with the ``tack" ion trap}
\author{G.Shu\footnote{School of Chemistry and Biochemistry, Georgia Institute of Technology, Atlanta GA 30332}}
\email{shugang@gatech.edu}
\author{C.-K. Chou}
\author{N.Kurz\footnote{Nion Company, Kirkland WA 98033}}
\author{M.R.Dietrich\footnote{Argonne National Laboratory, Lemont IL 60439}}
 \author{B.B.Blinov}
\affiliation{Department of Physics, University of Washington Seattle WA 98195}
\date{\today}

\begin{abstract}
Trapped, laser-cooled ions produce intense fluorescence. Detecting this fluorescence enables efficient measurement of quantum state of qubits based on trapped atoms. It is desirable to collect a large fraction of the photons to make the detection faster and more reliable. Additionally, efficient fluorescence collection can improve speed and fidelity of remote ion entanglement and quantum gates. Here we show a novel ion trap design that incorporates metallic spherical mirror as the integral part of the trap itself, being its RF electrode. The mirror geometry enables up to 35\% solid angle collection of trapped ion fluorescence; we measure a 25\% effective solid angle, likely limited by imperfections of the mirror surface. We also study properties of the images of single ions formed by the mirror and apply aberration correction. Owing to the simplicity of its design, this trap structure can be adapted for micro-fabrication and integration into more complex trap architectures.
\end{abstract}
 \maketitle
 
Trapped ion qubits are among the most promising candidates for a practical quantum computer~\cite{Divincenzo-Dogma-2001, Kielpinski-2002,ion-photon-review}. The qubits are realized by the long-lived atomic levels such as the ground state hyperfine levels~\cite{Blinov-Hperfine-2004} or the metastable excited states~\cite{CaSidebandCooling}. Laser and RF control techniques enable single- and multi-qubit operations with high fidelity~\cite{twoqubitgate-2003}. One of the crucial components of a quantum computation, the qubit state readout, is accomplished by the selective electron shelving technique where one of the qubit states produces strong fluorescence, while the other remains dark~\cite{YbQubit-2007, HighEfficientDetect}. Efficient collection of single photons emitted by trapped ions excited by short laser pulses is essential for generation of remote ion entanglement, quantum gates between distant ions and quantum state teleportation protocols~\cite{Cirac-QuantumStateTransfer-1997, ion-photon-entanglement, Duan-Repeater-2004,Duan-RemoteGate-2006, Chuang-Remotegate, Monroe-2009, Moehring-IonEntanglement}. These tasks require high efficiency of ion fluorescence collection; the optical mode quality is also essential for the remote ion entanglement, teleportations and quantum gate protocols. In this paper, we address this problem from the perspective of integrating optics into the trap structure. 

A cold, trapped ion interacting with a resonant laser radiation represents an almost ideal point source of light. Thus the task of collecting the ion\rq{}s fluorescence becomes the problem of a large solid angle imaging of a point source. Additional restrictions imposed by the trap structure and the vacuum system tend to limit the usable solid angle to $\sim$0.5~sr, or a few percent of the total solid angle. Traditional high numerical aperture (NA) microscopy setups are not applicable because of the long working distances involved in the trapped ion imaging. The multi-element refractive optics quickly becomes too bulky (and expensive) as the NA increases~\cite{Weiss-AtomImag-2007, blatt-IonMirrorImageInterference-2001}. The use of near field optics, such as~\cite{Greiner-2009} is generally not applicable to trapped ions due to charging of the dielectric surface of the optics~\cite{InchamberLens-2001}; a notable exception is the diffractive microlens used to produce diffraction-limited ion images in~\cite{diffractionLens-2011}. Concave mirrors~\cite{Leuchs-FreespaceModeConvert-2007} have been proposed as a way to potentially increase the collection solid angle to almost 90\%; such a deep parabolic mirror can be used in conjunction with the ``stylus" ion trap~\cite{Wineland-stylustrap-2009}. Another alternative is the use of optical cavities surrounding the trapped ion~\cite{Keller-IonCavity-2004, blatt-cavity-2009}, where a very large fraction of the ion\rq{}s spontaneous emission would go into the cavity mode due to the Purcell effect. All of these methods present many challenges, and are currently being actively pursued.

Following our success with utilizing a spherical mirror to collect approximately 10\% of fluorescence from an ion isolated in a standard linear RF trap~\cite{BigMirrorTrap-2009-1,shu2010}, we designed, built and tested a novel ion trap - the ``tack" trap (Fig. 1). The trap uses a concave conductive optical surface as one trap electrode and a sharp needle passing through the axis of the mirror as the other, thus resembling the ubiquitous office tool (Fig. 1(a)). Additional ring electrode is placed above the mirror surface to increase the trap depth. The rotational symmetry of the optical surface forms a trapping potential pattern co-aligned with the optical axis (Fig. 1(b)). This highly integrated approach has little optical blocking, therefore it can have a very large accessible solid angle solely restricted by the shape of optics. The needle electrode forms a very large local field gradient, resulting in a relatively deep trapping potential localized in a small volume (Fig 1(c,d)). The ion\rq{}s position can then be precisely controlled anywhere along the optical axis by moving the needle electrode. 

\begin{figure}[htp]
\centering
\includegraphics[width=0.7\textwidth]{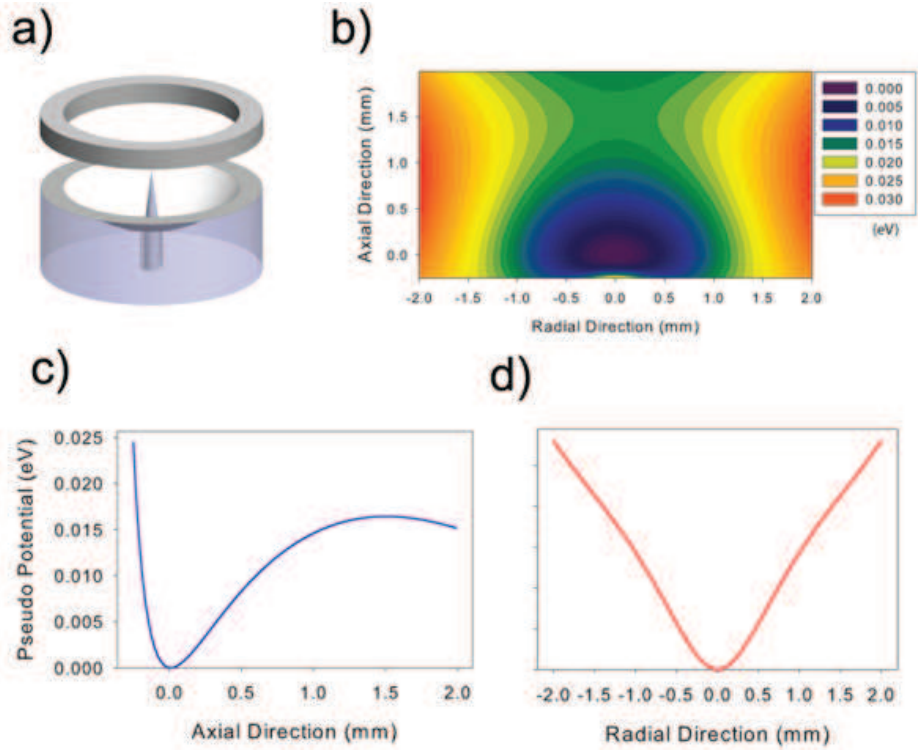}
\caption{Concept of the tack trap~(a) and pseudopotential simulation: near the trapping center~(b); along the trap axis of symmetry~(c), and along the radial direction~(d). The numbers are given in electron-volts for typical trapping RF parameters: RF frequency of 23~MHz and RF voltage of 270~V.}
\label{ConceptAndSimulation}
\end{figure}

\begin{figure}[htp]
\centering
\includegraphics[width=0.7\textwidth]{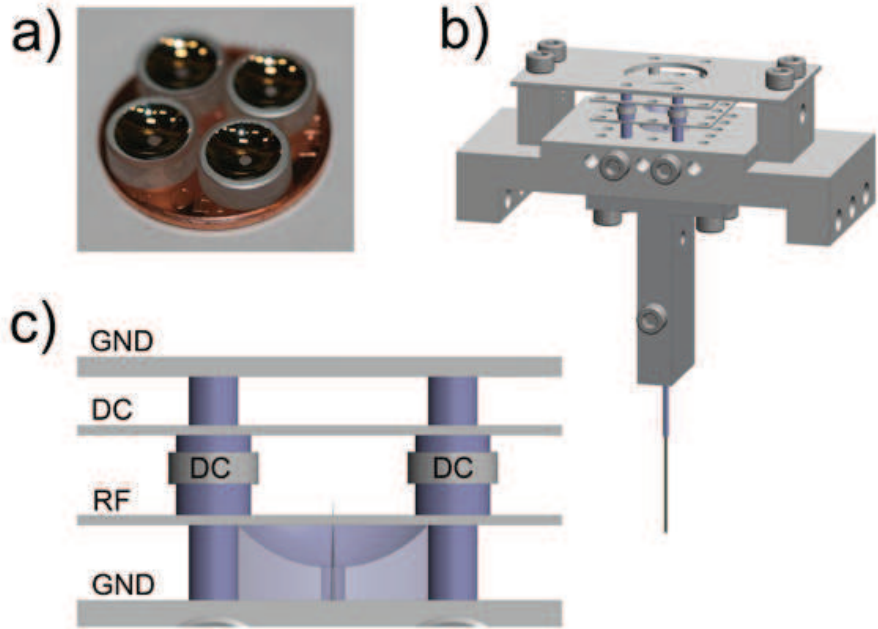}
\caption{The construction of the tack trap. Four spherical mirrors with holes drilled are placed on a U.S. penny for scale~(a). The overview of the trap structure showing the aluminum frame and structural other elements, as well as the needle and needle guide~(b). The side view showing the electric connection of the trap, with conductors shown in gray color and insulators in light blue~(c).}
\label{TrapBuilding}
\end{figure}

This trap design can be readily applied to any conductive optical surface shapes with rotational symmetry, including spherical, parabolic and elliptical mirrors. We tested the concept with a commercially available, low-cost spherical mirror. As shown in Fig.~\ref{TrapBuilding}, the trap is built around a 6~mm diameter, 4~mm radius of curvature Al-coated concave spherical mirror (Anchor Optics part number AX2740). A $\O0.75$~mm hole is drilled through the vertex of the mirror to accommodate a 0.5~mm diameter tungsten rod that tapers to a point. The rod is attached to a linear vacuum actuator feedthru driven by a precision micrometer. The trap assembly is supported by the superstructure made from Kimball Physics Inc. e-beam parts and a custom-made aluminum holder (Fig.2(b)). Four micromotion compensation electrodes are attached to the ceramic rods (Fig. 2(c)). The top plate may be biased, also for for micromotion compensation. The RF is applied to the metallic surface of the mirror, while the tungsten rod is RF-ground. 

According to our simulations, the pseudopotential of the tack trap, being symmetric around the optical axis, is stronger in the axial direction, and weaker along the radial. For a typical RF drive voltage of 270~V at 23~MHz, the trapping depth is on the order of 0.05~eV. This trapping depth is comparable with that of a surface trap~\cite{surfacetrap-2005, SandiaY-2011}, which one might expect from the topological similarity of these two trap types. However, the trap\rq{}s sharp needle and the curved metallic reflector surface increase its pseudopotential depth compared to that of a planar trap with similar dimensions. Our simulations also indicate, that as the needle is translated along the optical axis, the trapping depth remains approximately the same for a wide range of needle positions. This enables us to first load ions with the needle extended further out, at the location more accessible to atomic beam, and then move the ions to the optimal position for imaging and light collection. 

The tack trap is placed in a Kimball Physics 4~1/2 inch \lq\lq{}spherical octagon\rq\rq{} vacuum chamber and mounted with its axis parallel to the optical table. The trap is operated at the pressure of about $3\times10^{-11}$~torr in a standard ultrahigh vacuum setup with a 20~l/s ion pump and a titanium sublimation pump. We use $^{138}Ba^+$ to test our trap. The Doppler-cooling of the trapped ions is achieved by combining a frequency-doubled 986~nm external-cavity diode laser (ECDL) and 650~nm ECDL in a single mode optical fiber~\cite{Matt-2009}, and focusing the combined beam almost perpendicular to the optical axis of the trap (which is also one of the trap's principal axes) with a small tilting angle of 4 degree to be able cool the axial motion of the ions. The atomic barium beam is generated by an oven located to the side of the trap, with barium beam pointing to the optical axis. Barium atoms are ionized using a resonant two photon-transition using a 791~nm ECDL laser and a 337~nm Nitrogen gas laser~\cite{BaPhotonionization-2007}, both aligned perpendicular to the Ba atomic beam. A two-stage long working distance microscope is setup along the optical axis and a 50/50 beam splitter is used to both image the ions using an Andor iXon electron-multiplied charge coupled device (EMCCD) camera and to count photons using a Hamamatsu photomultiplier tube (PMT). The sharp needle tip, illuminated by the focused laser light, is used as a reference point to align and focus the optical system. 

The ions are trapped at a distance of 0.541~mm from the needle tip, consistent with our simulation prediction of 0.55~mm. At the background pressure of $3\times10^{-11}$~torr, for a single $^{138}Ba^+$ ion, we observed $>$30~minutes dark lifetime and $>$48~hours lifetime with laser cooling. We also loaded and laser-cooled various 2-dimensional ion crystals, as shown in Fig.~3. The 2-dimensional nature of these crystals is easily understood based on the strength and the symmetry of the trapping potential. This trap is in a way complimentary to the more typical linear RF ion trap, where the axial direction is the weakest, and elongated ion crystals are formed. The ``pancake" crystals formed in the tack trap were stably trapped for many hours. While micromotion cannot be nulled for the ions located off center in these crystals, this trap design may be potentially used as a platform for quantum simulations with trapped ions~\cite{IsingSimulation-2010}.
 
\begin{figure}[htp]
\centering
\includegraphics[width=0.7\textwidth]{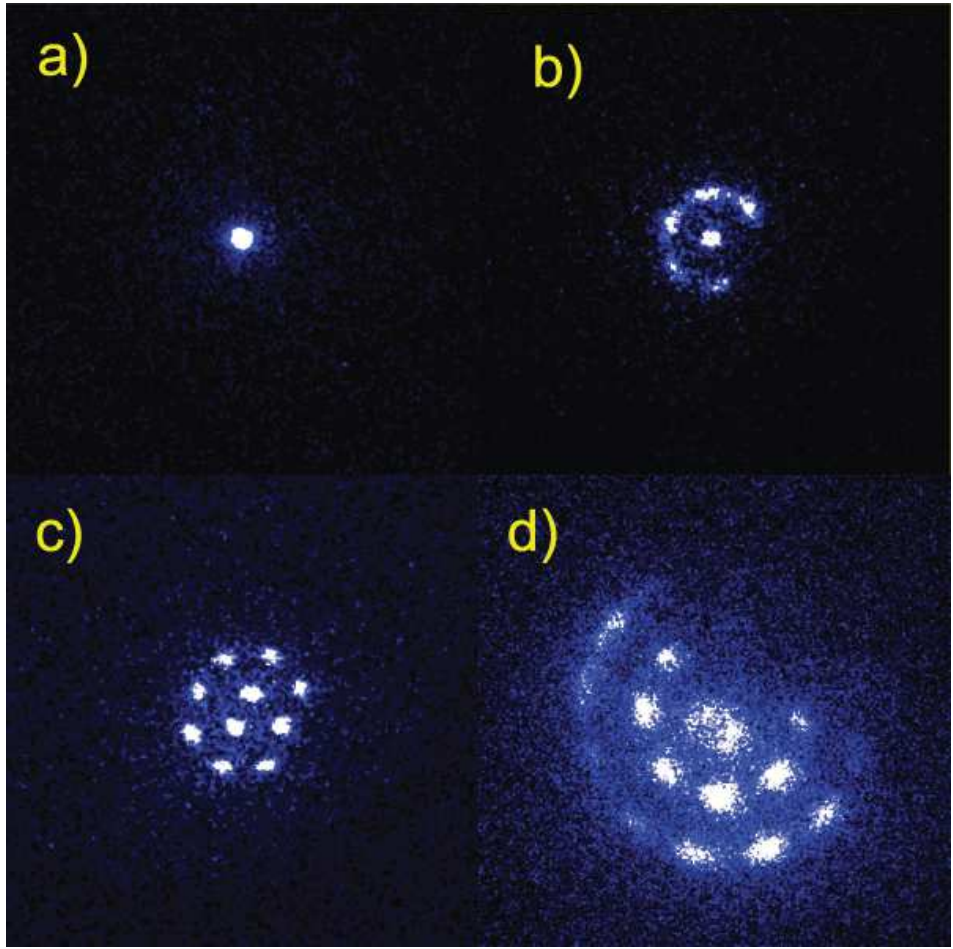}
\caption{Images of Ba$^{+}$ ions in the tack trap. A single ion~(a), a symmetric 7-ion crystal~(b), an asymmetric 10-ion crystal~(c) and a larger (24-ion) crystal~(d). All 10 ions are visible in (c), while (b) and (d) have some dark ions representing different isotopes of barium. Images in (a), (b) and (c) are directly formed by a microscope objective, while (d) is formed by the mirror.}
\label{ionpicture}
\end{figure}
  
By adjusting the needle electrode position, ions can be moved over a fairly large distance ($>1.5~mm$), covering the entire range from the mirror focus to its curvature center. We measured the position of the trapped ion relative to the reading on the micrometer translating the needle, and found a nearly linear dependence, which only starts to deviate from the straight line at the largest needle extensions. The needle can be reliably moved over the entire range without ion loss. To further characterize the trap, we measured the trap\rq{}s secular frequency by sending in a 10~V amplitude oscillating  voltage to one of the compensation electrodes and scanning its frequency from 100~kHz to 4~MHz. The dips in ion fluorescence correspond to the radial and the axial frequencies, which were found to be about 200~kHz and 420~kHz, respectively. A closer look at the radial frequency reveals small non-degeneracy in the radial direction, which is probably due to slight misalignment between the needle and the main optical axis.

\begin{figure}[htp]
\centering
\includegraphics[width=0.7\textwidth]{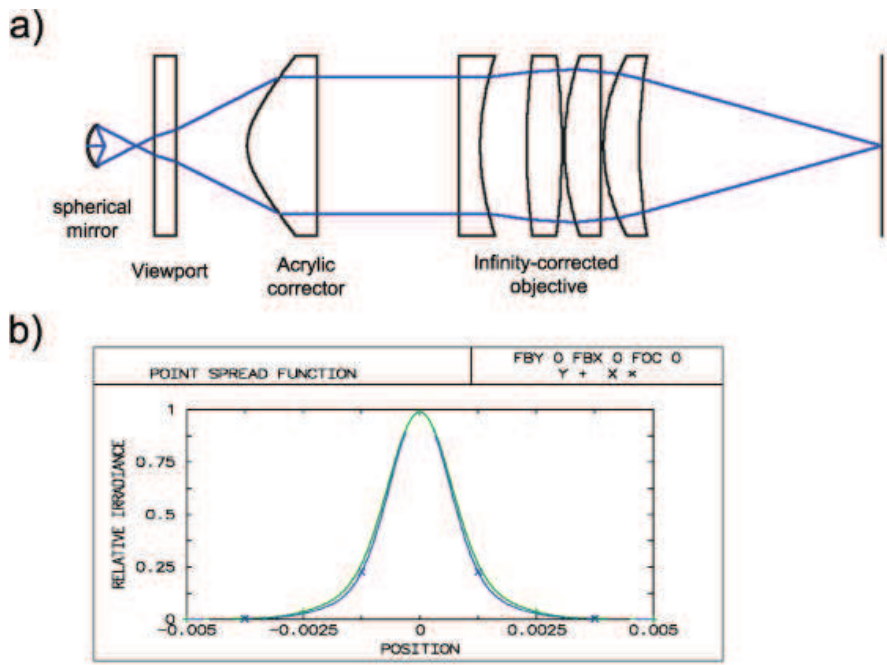}
\caption{The aberration compensation scheme. The schematics of the optical setup showing the mirror, the vacuum viewport, the aspheric element, and the multielement low-NA micro objective~(a). The ideal point spread function calculated from simulations~(0.97~$\mu m$) reaches diffraction lilmit~(1.15~$\mu m$)~(b).}
\label{compensation}
\end{figure}

Our mirror electrode covers approximately 38.6\% of the solid angle around its focal point, which is defined to be exactly halfway between the vertex of the mirror and its curvature center. But for such a high focal ratio system, the huge spherical aberration make the definition of the focal point rather useless. Additionally, due to the small size and high curvature of the mirror, the light reflected from it can not be compensated with a Schmidt-like corrector that we used in our earlier designs~\cite{shu2010}, since the rays reflected from different points in the mirror will have intersected by the time they reach the outside of the vacuum chamber. Therefore, we designed a different type of aspheric lens, such that it collimates the light from the ion while compensating its spherical aberration. The setup is shown schematically in Fig.~\ref{compensation}(a), and the expected diffraction-limited image point spread function in Fig.~\ref{compensation}(b). For best compensation results, instead of being at the focus of the mirror, the ion is moved 0.25~mm further away from the mirror. We numerically calculate the shape of the corrector using algorithm similar to to that described in~\cite{shu2010} and arrive at a high order nonlinear curve. 

\begin{figure}[htp]
\centering
\includegraphics[width=0.7\textwidth]{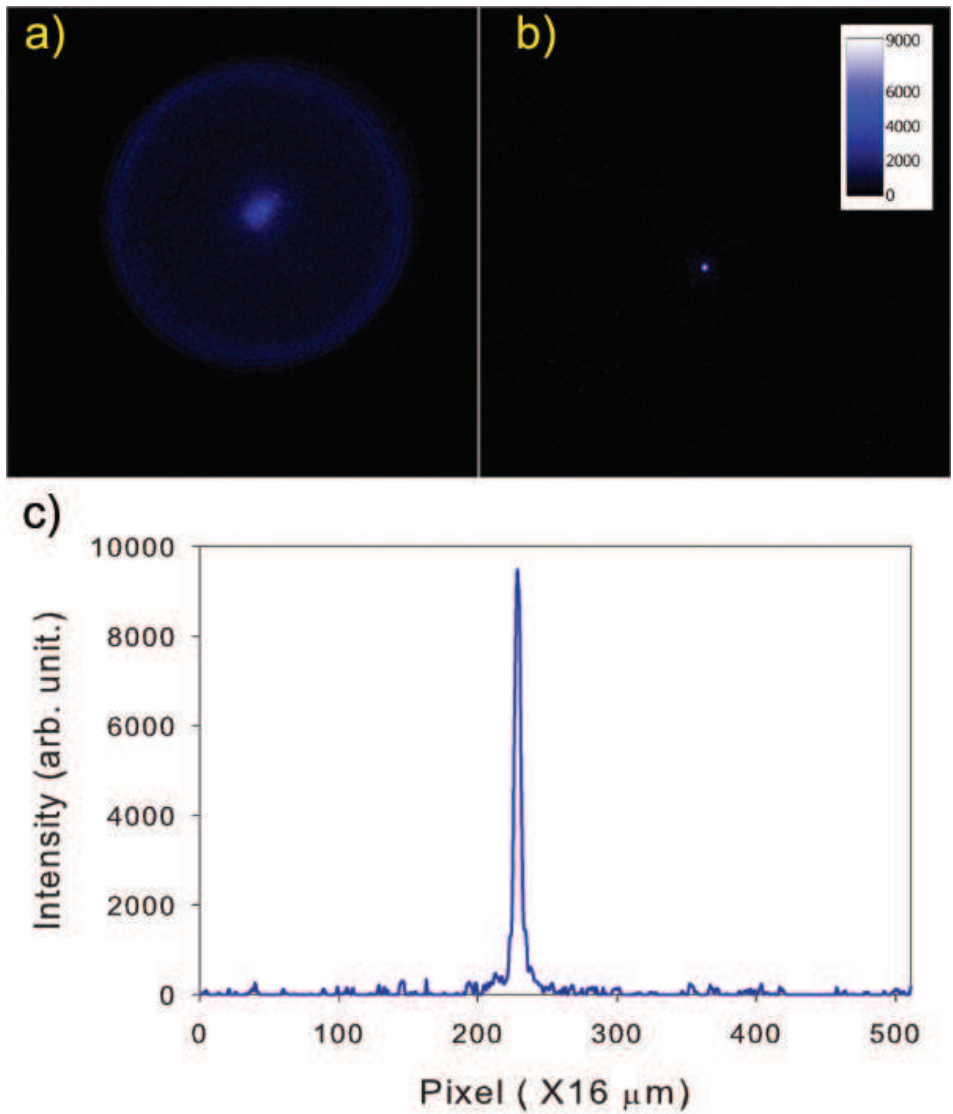}
\caption{Demonstration of the effectiveness of the optical compensation scheme. An ion crystal image without correction~(a), with large rings due to spherical aberration, where $\sim$90\% of the light from the ions is dispersed. A single ion image formed by the mirror-corrector system~(b), and the profile of the image~(c). The full width at half maximum of the peak is less than $100\mu m$, which is a significant improvement, but still far from diffraction limit ($1.15~\mu m$ for this setup). }
\label{performance}
\end{figure}

To test the effectiveness of this compensation scheme, we machine the aspheric lens prototypes out of clear acrylic. The obvious improvement of the image quality can be seen in Fig.~\ref{performance}. Here, in Fig~\ref{performance}(a), an uncompensated image of an ion crystal is shown, with the bright dots in the center formed by the paraxial rays from the mirror, while much of the light is lost into the large, diffuse rings. In Fig.~\ref{performance}(b), a single ion image with the compensation scheme is shown. Although the aberration is greatly reduced, it is obviously not fully compensated with our current device. This is largely because both the spherical mirror and the aspherical lens are far from perfect. Notably, a direct measurement of the mirror surface with a measuring microscope~(OGP SmartScope) shows that its radius of curvature is at least $\sim10\%$ larger than its nominal value; the surface is not a true sphere either. The precise control and measurement of the plastic home-made corrector surface is even more difficult. Taking these factors into account, we believe that our compensation scheme performs exceptionally well, and we clearly demonstrate the proof-of-principle.

Geometrically, this spherical mirror covers about 35\% of the solid angle from the ion placed 0.25~mm away from the focus. To calibrate the mirror\rq{}s efficiency for light collection, we use single photon generated by repeated shelving and deshelving of the single ion, using the scheme described in ~\cite{shu2010}. In the test we counted $9957\pm 611$ photons per 1,000,000 excitations of the ion, corresponding to 1\% overall photon collection and detection efficiency. After factoring out the quantum efficiency of the PMT, and the transmissive and reflective losses of all optical elements, we arrive at approximately 24\% of the entire 4$\pi$ solid angle to be subtended by the mirror, which is equivalent to about 3.0~sr, or a numerical aperture of 0.85. There are several possible reasons for the reduced efficiency, including the non-uniform spatial distribution of the ion dipole radiation pattern, and the degradation of the unprotected aluminum surface of the mirror during the baking of the vacuum system~\cite{AlSurface-1996}. 
 
Our trap design can be readily applied to other shapes of reflectors, such as parabolic or elliptical mirrors.  Parabolic mirrors can in theory reach solid angle close to 100\%~\cite{Leuchs-FreespaceModeConvert-2007} and produce diffraction-limited, collimated light from an ion located in the focus of the mirror. A proposal exists to combine such a mirror with the stylus trap designed by the NIST group~\cite{Wineland-stylustrap-2009}. We believe that a simpler design similar to that described here may be more practical. It does not require micro fabrication and, with a mirror surface being the RF electrode, the tack trap can sustain higher RF voltages than the stylus trap, thus improving the trapping depth. 

Elliptical mirrors are also interesting, since they can produce, in one of their foci, diffraction-limited image of a point source located in the other focus. Thus, ions placed in one focus point can be directly imaged into a single mode optical fiber located at the other focus point without additional optics. For free space optics, elliptical mirrors can be used as a numerical aperture converter by  reducing a large-NA at one focus into a smaller-NA one at the other; the successive imaging can then  be done with a low-NA microscopy setup. Additionally, elliptical mirrors appear to be suitable for microfabrication and integration into state-of-the-art chip-scale ion traps, as discussed below.

An important criterion of an ion trap design, especially in quantum information applications, is its feasibility for scaling - both down in size and up in number. The simplicity of the tack trap permits large scale fabrication and integration with the microfabricated planar traps and trap arrays.  One challenge here is to fabricate the highly curved surface controllably. There are efforts currently under way to integrate curved mirrors into a planar ion traps using more traditional methods~\cite{MEMS-ASpherical-Vdovin-2003, TrueMirror-2011}, but it is challenging to produce a large focal ratio optical surface while controlling its exact shape. We propose a self-assembly molding procedure, with which a full control of the shape and quality of optical surface is possible~\cite{MEMS-BulkSiMirror-2003, MEMS-MicroMirror-Bartlett-2003}. 

\begin{figure}[htp]
\centering
\includegraphics[width=0.7\textwidth]{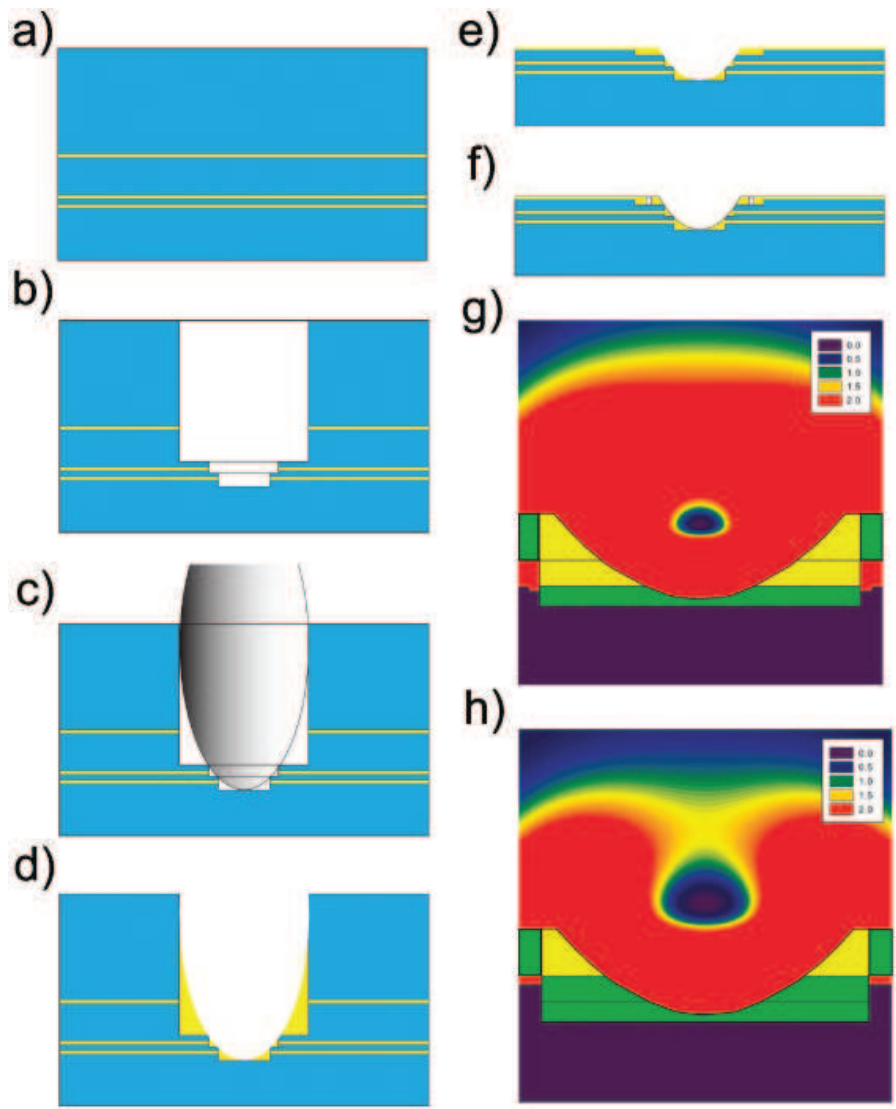}
\caption{A proposal for microfabricaton of mirror traps. First, a substrate with appropriate conductive layers is prepared~(a). A series of step guides are etched with standard microelectromechanical systems (MEMS) techniques such as DRIE~(b). An ellipsoid bead falls into the guide, finishing a self-assembly~(c). Current is run through all the conductive layers to plate metal into the gap between the bead and the substrate; the bead is then removed with chemical dissolution~(d). The substrate can then be polished down to the appropriate height~(e). The mirror surface can then be segmented and electrical connections are made~(f). Two pseudopotential simulations show how the height of the trapping zone can be controlled by routing RF and ground electrodes to different parts of the mirror, where RF is represented by yellow and ground by green. When the RF is applied on the upper two segments of the mirror, the trapping zone is lower~(g), while if the RF is applied to the top segment only, the trapping zone is raised higher, close to the focus of the elliptical mirror~(h). The simulations predict deeper trapping potentials than in a planar trap of a similar scale.}
\label{MEMS}
\end{figure}

Micron-scale spherical or ellipsoid beads with optical quality surface are available from either commercial sources, or by in-situ lab fabrication (e.g. by melting glass fiber into a microsphere with laser blasting~\cite{LaserBlastingFiberCavity-2000}). With single point diamond turning technology, the mold or even the surface itself can be directly machined with ultra high precision~\cite{rohtua-micromachining-2003}. The concave reflector surface can then be fabricated by electroplating high-reflectivity metal, such as aluminum or gold, with the bead serving as a mold. The possible fabrication steps are outlined in Fig.~\ref{MEMS}. After the substrate is prepared with series of conducting and insulating layers, a guiding structure is etched to ensure alignment of the electroplated mirror with the other parts of the trap. Then the substrate is patterned with a thin conductive layer for electroplating. When the beads and the substrate are put together (typically inside some liquid), the beads will fall into the guides due to gravity, thus finishing the self-assembly. During the electroplating, the metal fills the gap between the bead and substrate (Fig.~\ref{MEMS}(d)). The beads can then be removed by chemical dissolution that would keep the metal surface intact. The finished metal surfaces will have the exact shape and smoothness of the molds. The fabrication of the needle electrode can be adapted to different trap scales. For example, for a trap with the characteristic structure size of $100\mu m$ or above, the needle can be fabricated separately by the deep-reactive ion etching (DRIE) process~\cite{MEMS-Needle-Bohringer-2003} and precisely aligned and bound to the mirror substrate. For smaller trap sizes, the needle can be eliminated altogether by dividing the mirror surface into different segments acting as the RF and the ground electrodes, as shown in Fig.~\ref{MEMS}(g) and (h).

In summary, we present a novel ion trap design for efficient fluorescence collection from trapped ions. We achieve an effective numerical aperture of 0.85, which, to our knowledge, is by far the largest solid angle for the ion traps. We discuss and test the possibility of aberrations compensation, and propose practical approaches for scaling down and integrating this ``tack" trap with micofabricated planar trapping system using the available MEMS technology. This trap design can be used in many trapped ion quantum computation and information, as well as for quantum simulations.

This research was supported by the National Science Foundation Grants No.~$0758025$ and No.~ $0904004$, the Army Research Office, and the University of Washington Royalty Research Fund.

%

\end{document}